# Chapter 1

# Quantum Serverless Paradigm and Application Development using the QFaaS Framework

**Hoa T. Nguyen[1], Bui Binh An Pham[1], Muhammad Usman[2,3], and Rajkumar Buyya[1]**

[1]*Cloud Computing and Distributed Systems (CLOUDS) Laboratory, School of Computing and Information Systems, The University of Melbourne, Parkville, 3052, Victoria, Australia*

[2]*School of Physics, The University of Melbourne, Parkville, 3052, Victoria, Australia*

[3]*Data61, CSIRO, Clayton, 3168, Victoria, Australia*

## 1.1 Introduction

Quantum computing promises to solve problems that are currently intractable for classical computers. Potential applications span various domains, including finance optimization problems [2], drug discovery [3], and complex system simulations [4]. However, the practical utilization of quantum computers remains a significant challenge due to the early stage of quantum software engineering [5] and the limitations of contemporary quantum hardware, known as the Noisy Intermediate-Scale Quantum (NISQ) devices [6]. These devices are characterized by limited qubits and susceptibility to various noises, which constrains their computational capabilities and the accuracy of the quantum execution results.

In this context, the emergence of the serverless computing paradigm offers a promising pathway to harness quantum computing more effectively. Serverless computing, mainly through the Function-as-a-Service (FaaS) model, allows developers to focus on writing code without the need to manage the underlying infrastructure [7]. This model has proven successful in classical computing by enhancing scalability, reducing operational overhead, and enabling rapid deployment of microservices [8]. Extending this paradigm to quantum computing, the Quantum Function-as-a-Service (QFaaS) framework [1] aims to provide similar benefits by abstracting the complexities of quantum hardware and seamlessly integrating quantum functions into hybrid quantum-classical applications.

The QFaaS framework is designed to facilitate the development and deployment of quantum applications by leveraging the serverless model. It addresses the critical challenges in quantum software engineering, such as the diversity of quantum programming languages and the efficient integration of quantum and classical computational resources [9]. By supporting multiple quantum software development kits (SDKs) like Qiskit, Q#, Cirq, and Braket, QFaaS offers a unified platform that mitigates vendor lock-in, enhancing the portability of quantum applications across different quantum cloud providers.

QFaaS incorporates several innovative features to streamline quantum application development. It provides an adaptive backend selection policy to automatically select the most suitable quantum processor based on the characteristics of the quantum circuit and the current workload of available quantum devices. This policy is crucial for optimizing the execution of quantum functions, considering the variability in performance and availability of NISQ devices. Additionally, QFaaS supports the development of service-oriented quantum applications [10] by providing essential DevOps capabilities, such as continuous integration and continuous deployment (CI/CD). This integration ensures that quantum functions are reliably deployed, monitored, and scaled according to user demand, facilitating quantum research transition from theoretical exploration to practical implementation. This chapter uses QFaaS as a practical example to discuss the serverless quantum computing paradigm and details how to set up and use QFaaS for deploying service-oriented quantum applications.

The rest of the chapter is organized as follows: It discusses the emerging paradigm of serverless quantum computing, specifically focusing on the QFaaS framework example. Our previous work [1] provides details of the architecture design and techniques used in the QFaaS framework. The chapter then describes the deployment details of QFaaS components and demonstrates examples of developing and deploying service-oriented quantum applications within this framework to realize the concept of serverless quantum computing. Finally, it discusses lessons learned, current limitations, and future work of QFaaS.

## 1.2 Serverless Quantum Computing

The serverless quantum computing paradigm represents a significant shift in how quantum computing resources are utilized and managed. By abstracting the complexities of infrastructure and resource management, serverless quantum computing allows developers to focus solely on the logic and functionality of their quantum applications [1]. This approach leverages the principles of serverless computing, such as Function-as-a-Service (FaaS), which have been successful in classical computing and extends them to the realm of quantum computing [7]. Serverless quantum computing, often called quantum serverless, simplifies the deployment and execution of quantum applications [11]. In a traditional computing model, developers must manage the underlying infrastructure, including servers, storage, and networking. This management is complex and resource-intensive, especially in the quantum domain, where specialized environments are required to host and operate quantum computers. Serverless computing removes these burdens by allowing cloud providers to handle the infrastructure management. Developers can focus only on deploying quantum codes triggered by specific events, and the cloud provider automatically manages the execution, scaling, and resource allocation [1].

This paradigm brings several advantages to quantum software engineering, including:

- **Abstraction of Quantum Infrastructure:** Serverless quantum computing abstracts the complexities of quantum hardware management [7]. Developers do not need to worry about quantum software environmental setup, qubit initialization, or quantum system operations. Instead, they can focus on developing quantum applications.

- **Scalability**: The serverless model inherently supports scalability [12]. Functions can be automatically scaled up or down based on demand. This dynamic scaling is particularly useful given the variability in quantum computing workloads. However, it is important to highlight that this advantage pertains to the scalability of classical infrastructure supporting deploying quantum functions, not quantum computer resources, due to the current early stage of quantum hardware development [1].

- **Cost Efficiency**: Serverless quantum computing operates on a pay-as-you-go model [8]. Users are billed based on the actual execution time and resources consumed by their quantum functions rather than paying for idle infrastructure. This model makes quantum computing more accessible and cost-effective for researchers and developers.

- **Integration with Classical Computing:** The serverless paradigm facilitates the integration of quantum and classical computing resources. Hybrid quantum-classical applications can be developed more quickly, leveraging the strengths of both quantum and classical processors [13].

Serverless quantum computing opens numerous possibilities for practical quantum applications [14]. For instance, quantum random number generation (QRNG) [15], [16] can benefit from the scalability and efficiency of the serverless model, providing truly random numbers for cryptographic applications. By abstracting the complexities of quantum hardware, offering scalable and cost-effective solutions, and facilitating the integration of quantum and classical computing resources, serverless quantum computing is poised to drive significant advancements in the field. As quantum hardware continues to evolve, the serverless model will play a crucial role in making quantum computing more accessible and practical for real-world applications in the future quantum utility era.

## 1.3 QFaaS Framework and Deployment

### *1.3.1 Overview of the QFaaS Framework*

The QFaaS framework is a comprehensive example of the serverless quantum computing paradigm [1]. The framework supports the development of service-oriented quantum applications, enabling researchers and developers to build and deploy quantum services without the need for manual environment setup and quantum hardware management.

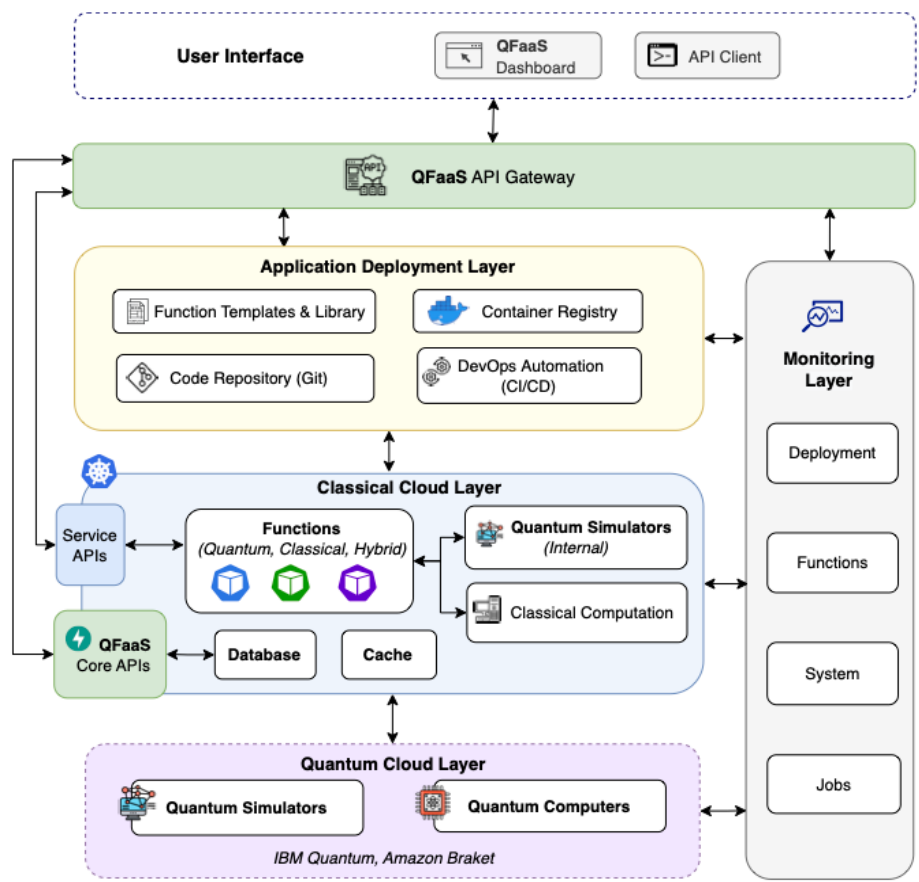

*Figure 1. A layered architecture of the QFaaS framework (Reused with permission from Nguyen et al. [1])*

As shown in Figure 1, the QFaaS framework comprises six key components:

1. **QFaaS Core APIs and the API Gateway** ensure secure interactions across all other components and manage all framework functionalities.
2. **Application Deployment Layer** manages function code version control, containerization, and function deployment automation with continuous integration and deployment support.
3. **Classical Cloud Layer** executes classical computation tasks, hosts the deployment of quantum functions as a Kubernetes deployment, and stores job execution results within the database.
4. **Quantum Cloud Layer** refers to external quantum providers that provide access to execute quantum tasks to external quantum simulators and quantum computers.
5. **Monitoring Layer** provides insights into system status and classical cloud resource consumption using monitoring tools, including Prometheus and Grafana.
6. **QFaaS UI** is a web-based user interface for developing, deploying, invoking, and monitoring quantum functions via interaction with the QFaaS Core APIs.

Figure 2 illustrates the flow of control from the creation of quantum functions to execution management. There are two types of users in QFaaS: quantum software engineers (developers) and end-users. The developer interacts with the framework to develop and deploy their quantum function codes as services and publish corresponding API endpoints. The function deployment is facilitated with continuous integration and continuous deployment (CI/CD) by the Application Deployment layer of the framework. End-users can then use the endpoint API of deployed quantum functions for invocation and integrate these API endpoints into their existing application workflow.

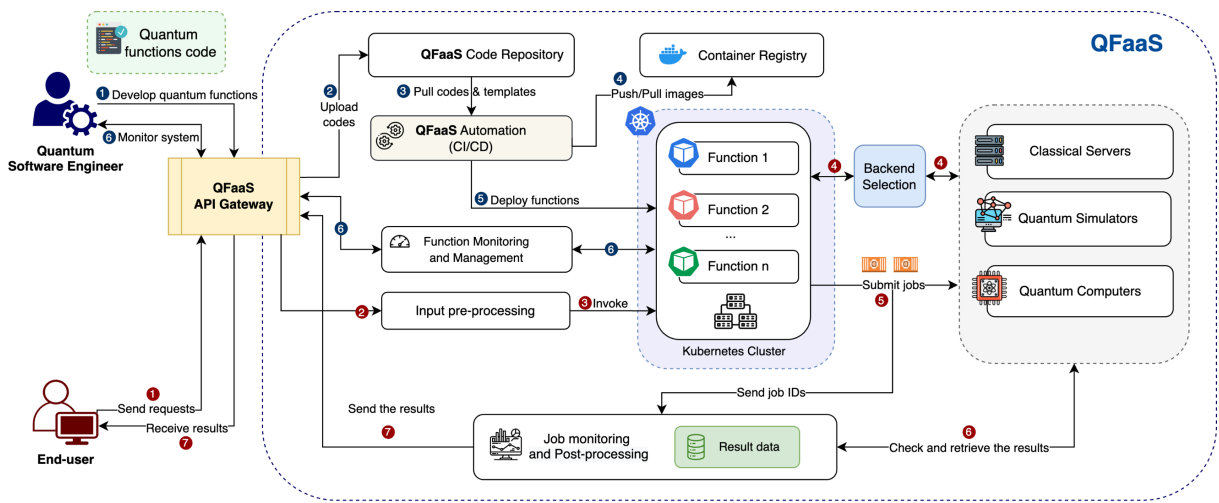

*Figure 2. A flow diagram of QFaaS operation*

For the sample deployment discussed in this chapter, we employed Kubernetes to orchestrate these resources. We used five Virtual Machines (VMs) running Ubuntu Linux. Four VMs were dedicated to running a Kubernetes cluster, with one master node and three worker nodes, while one VM served as the host for the Gitlab server, as shown in Figure 3.

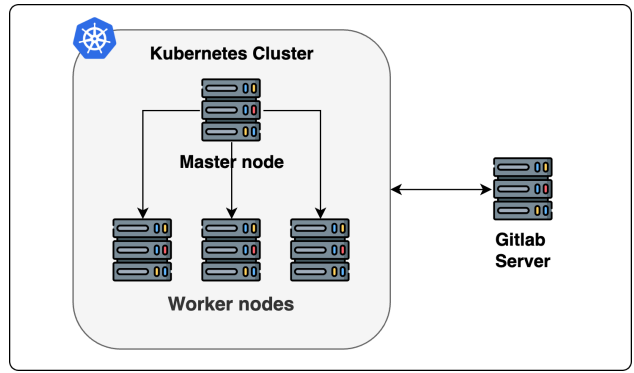

*Figure 3: Sample classical server setup for the deployment of the QFaaS framework*

The setup process starts with configuring a Git-based platform, which is essential for the function deployment component. Then, we set up a Kubernetes cluster on the classical cloud layer and supplementary components, such as the MongoDB database. Finally, we set up the QFaaS Core APIs and the UI to complete the setup process.

## *1.3.2 Setup QFaaS Application Deployment Layer*

In this section, we describe the Application Deployment Layer's set-up process, comprising a Git-based platform and a Docker hub repository. These components will manage the source code for quantum functions and their built images, respectively.

**Step 1: Setting up a Git-based platform.**

We use GitLab as it offers a self-hosted version, which is preferable if we want to deploy the whole QFaaS framework in a private cluster. For detailed instructions on installing Gitlab on Ubuntu, please refer to their official documentation at https://about.gitlab.com/install/#ubuntu.

After Gitlab is deployed, we need to create a new Git repository based on the template in the qfaas-fn folder from the repository at https://github.com/Cloudslab/qfaas. The Deployment Application Layer uses this repo to store the developed quantum functions source code. We need to register a new user in the self-hosted platform. The QFaaS framework will use this user credential to authenticate, access, and store users' code in the repository created earlier.

**Step 2: Create a Docker Hub repository.**

We can refer to the Docker Hub documentation https://docs.docker.com/docker-hub/repos/create/ for instructions on how to create a new repository.

**Step 3: Setting up a CI/CD pipeline (optional)**

We can establish a continuous integration and deployment (CI/CD) process to automatically deploy quantum functions whenever users update the source code. A script called `build.sh` is available in the template git repository at directory `functions/build.sh`. This script reads a `build.txt` file that contains a list of quantum functions, each separated by a new line. For each function, the script rebuilds its Docker image, publishes it to DockerHub, and then deploys it on Kubernetes using `faas-cli`.

The CI/CD configuration script that incorporates this process could resemble the following:

```
stages:
  - build

build:
  stage: build
  image: docker:23.0.3
  script:
    # [installing OpenFaaS CLI steps truncated]
    ...
    - cd functions
    - sh build.sh build.txt
    - docker image prune -a --force --filter "until=48h"
```

## *1.3.3 Setup QFaaS Classical Cloud components*

This section explains how to establish a new Kubernetes cluster and then deploy and initialize the MongoDB database.

**Step 1: Setting up a Kubernetes cluster.**

We are using MicroK8s to manage the Kubernetes cluster. It integrates the OpenFaaS framework [17], which is used as a base to extend with quantum serverless features.

1.1. Install `microk8s` on the Master node and three Worker nodes.

```
sudo snap install microk8s --classic
```

1.2. Connect three Worker nodes with the Master node.

For more information on how to do this, we can refer directly to `microk8s` official doc https://microk8s.io/docs/clustering

1.3. Enable OpenFaaS on the Master node.

```
microk8s enable community
microk8s enable openfaas
```

To verify that OpenFaaS is installed correctly, we can run

```
microk8s kubectl -n openfaas get deployments -l "release=openfaas,
app=openfaas"
```

An error may occur when executing the microk8s kubectl command for the first time because the default Ubuntu user lacks sufficient permissions to access MicroK8s resources. This issue can be resolved by re-running the command with `sudo` or by adding the Ubuntu user to the MicroK8s group as follows:

```
sudo usermod -a -G microk8s ubuntu
sudo chown -R ubuntu ~/.kube
newgrp microk8s
```

1.4. Install `openfaas` CLI.

```
curl -sSL <https://cli.openfaas.com> | sudo -E sh
```

1.5. Create `qfaas` Kubernetes namespace, which would group all deployed functions.

```
microk8s kubectl create namespace qfaas
```

**Step 2: Deploy the MongoDB database.**

The database stores and manages user information, permissions, and metadata from completed jobs and quantum functions, including execution time, and job results. This section details the process of deploying a MongoDB database onto the cluster, followed by the addition of simulated user data. This setup will facilitate subsequent login and testing of the QFaaS framework.

1. Clone the QFaaS source code repository

   ```
   cd ~
   git clone https://github.com/Cloudslab/qfaas
   ```

2. Navigate to the `mongodb` directory containing the Kubernetes deployment scripts for setting up the MongoDB database.

   ```
   cd qfaas/mongodb
   ```

3. Manually reserve persistent volumes for MongoDB

   ```
   # create persistent volume (PV)
   ```

```
microk8s kubectl apply -f pv.yaml
# create persistent volumn claim (PVC)
microk8s kubectl apply -f pvc.yaml
```

4. Install MongoDB

    Here is what the `deployment.yaml` file looks like. We need to provide the credentials for the root MongoDB user, which QFaaS Core APIs will use to read and write to the database.

    ```
    # deployment.yaml
    apiVersion: apps/v1
    kind: Deployment
    metadata:
      labels:
        app: mongodb
      name: mongodb
      namespace: qfaas
    spec:
      replicas: 1
      selector:
        matchLabels:
          app: mongodb
      template:
        metadata:
          labels:
            app: mongodb
        spec:
          containers:
          - name: mongodb
            volumeMounts:
              - mountPath: /data/db
                name: mo-data
            image: mongo:4.4
            ports:
            - containerPort: 27017
            env:
              - name: MONGO_INITDB_ROOT_USERNAME
                value: <username>
              - name: MONGO_INITDB_ROOT_PASSWORD
                value: <password>
          volumes:
          - name: mo-data
            persistentVolumeClaim:
              claimName: mo-data-pvc
          restartPolicy: Always
    ```

    Then, run the following command to apply the script, creating a pod running MongoDB under the `qfaas` namespace.

    ```
    microk8s kubectl apply -f deployment.yaml
    ```

5. Initialize the required collections in the MongoDB database and add admin user data.

    ```
    cd ~/qfaas/mongodb
    python dbInit.py
    ```

## 1.3.4 QFaaS Core APIs

The QFaaS Core APIs serve as the backbone of the QFaaS Framework, offering essential services such as access control, management of quantum functions, execution of quantum jobs, secure connections to external quantum computing resources, and selection of the most suitable quantum backend. These functions collectively support quantum functions' straightforward development, deployment, and running. Figure 4 illustrates the overview of the class diagram of QFaaS Core APIs.

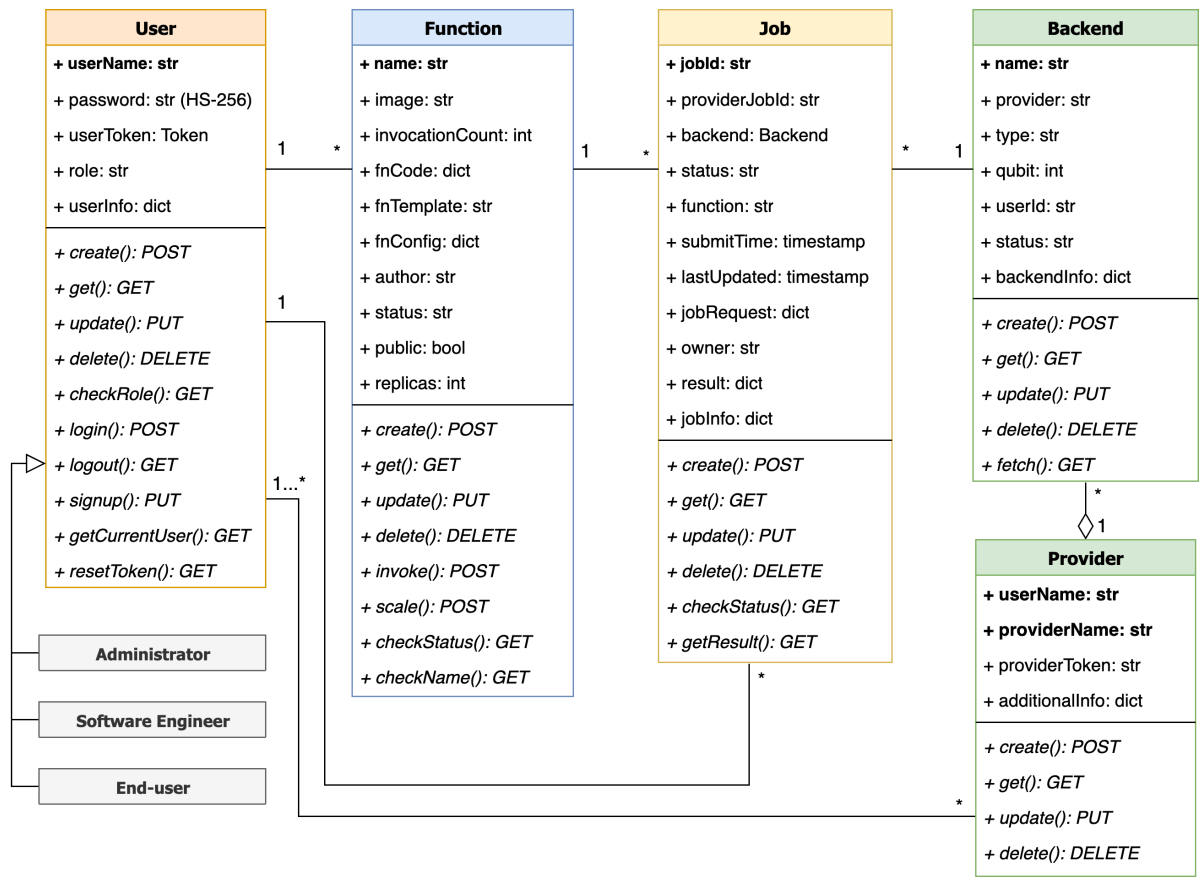

*Figure 4. Overview of QFaaS Core API components*

The rest of this section describes the deployment process of QFaaS Core APIs.

**Step 1: Configure the `.env` file.**

1. Start by duplicating the `.env.example` file in the `qfaas` folder we have cloned previously.

   ```
   cd ~/qfaas
   cp .env.example .env
   ```

2. The file `.env` should look like the following:

```
## 1. MongoDB
MONGO_DETAILS='mongodb://<username>:<password>@mongodb.qfaas.svc.clus
ter.local:27017/'

## 2. Gitlab
GIT_BRANCH=
QFAAS_USER_GITLAB=<username>
QFAAS_EMAIL_GITLAB=<user_email>
QFAAS_PASSWORD_GITLAB=

ROOT_PATH='/app/'

## 3. DockerHub
DOCKER_REPOSITORY=

## 4. QFaaS
QFAAS_URL="http://<master-node-ip-address>::<port>"
QFAAS_USER=admin
QFAAS_PASSWORD=
QFAAS_FUNCTION_URL=.openfaas-fn.svc.cluster.local:8080
```

3. We must provide the `username` and `password` we previously set when installing the MongoDB server.
4. `GIT_BRANCH` is the main branch of the Functions git repository. `QFAAS_USER_GITLAB`, `QFAAS_EMAIL_GITLAB` and `QFAAS_PASSWORD_GITLAB` are the credentials of the Gitlab user.
5. `DOCKER_REPOSITORY` is the name of our DockerHub repository, which will be used to store the built images.
6. For `QFAAS_URL`, we can get the port of the Gateway by running:

```
  microk8s kubectl get svc -o wide gateway-external -n openfaas
```

`QFAAS_USER` is set to `admin` by default. And `QFAAS_PASSWORK` should be set to

```
echo $(microk8s kubectl -n openfaas get secret basic-auth -o
jsonpath="{.data.basic-auth-password}" | base64 --decode)
```

**Step 3: Create Kubernetes secret based on the `.env` file**

```
microk8s kubectl create secret generic env-secret --from-env-file=.env -
n=qfaas
```

This command creates a new secret called `env-secret` under the `qfaas` namespace with all the keys and values from the `.env` file.

**Step 4: Install QFaaS Core**

```
cd ~/qfaas/kubenetes
kubectl apply -f deployment.yaml
```

**Step 5: Expose port 5000 of QFaaS Core API to make it accessible from the Internet.**

```
microk8s kubectl apply -f ~/qfass/kubenetes/service.yaml
```

## *1.3.5 Deploy QFaaS UI*

QFaaS UI is a web-based user interface developed using React to interact with QFaaS Core APIs to facilitate end users' usage instead of interacting directly with the APIs set. This step is optional but can provide users with an interactive way to utilize the functionalities of the QFaaS framework. The sample deployment of QFaaS UI is described below:

Before the deployment, we can customize the QFaaS UI source code and its Dockerfile to create a Docker image that contains the latest source code. Then, we can set up QFaaS web UI using `deployment.yaml` and `service.yaml` files provided, similar to deploying QFaaS Core APIs, to create its corresponding Kubernetes deployment and service instance. We also need to configure the Kubernetes Ingress if we need to bind a domain name with the QFaaS UI service. The sample Ingress configuration `ingress.yaml` is also provided in our package. For example, we map the domain name `qfaas.cloud` to the QFaaS UI service as follows:

```yaml
apiVersion: networking.k8s.io/v1
kind: Ingress
metadata:
  annotations:
    nginx.ingress.kubernetes.io/auth-type: basic
    nginx.ingress.kubernetes.io/auth-secret: basic-auth
    nginx.ingress.kubernetes.io/auth-realm: 'Authentication Required - foo'
  name: qfaas-ui-ingress
  namespace: qfaas
spec:
  ingressClassName: nginx
  rules:
  - host: <domain name>
    http:
      paths:
        - path: /
          pathType: Prefix
          backend:
            service:
              name: qfaas-ui-service
              port:
                number: 80
                apiVersion: networking.k8s.io/v1
  tls:
  [extra HTTPS configuration]
```

Once the deployment is finished, we can log in to the web-based UI at the `<domain name>` (e.g., qfaas.cloud) using the default Admin user credentials we established earlier (as shown in Figure 5). We advise users to change the default password after the initial login.

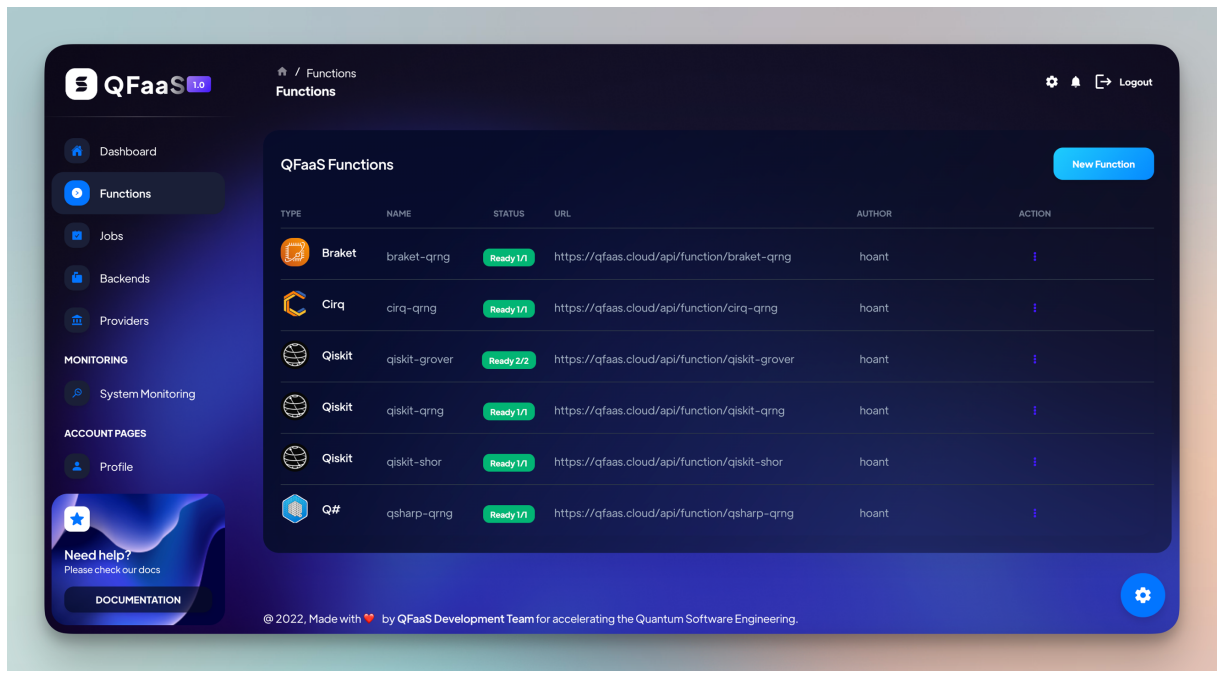

Figure 5. QFaaS UI page for managing all quantum functions

## 1.4   Service-oriented Quantum Application Deployment

This section describes the sample process of developing, deploying, and using sample quantum functions in QFaaS [1].

### *1.4.1 Quantum Functions and Quantum Services*

A **quantum function** in QFaaS can be considered as a unit of a quantum application, which can contain single or multiple functions. We note that not all quantum applications are suitable for deployment as a service, akin to the function and service counterparts in classical serverless computing [1]. The best practice we suggested for creating a quantum function as a service is as follows:

- Each quantum function has a single responsibility, with the main functionality implemented by a single quantum circuit. This approach makes the function easier to maintain and scale.

- Ensure functions are stateless, without relying on the underlying infrastructure to hold state between executions. We can use the database to manage the state and execution output of the quantum function.
- The circuit should be adjustable based on inputs of the user invocation, not a fixed quantum circuit.
- We are keeping functions lightweight to mitigate cold start times by minimizing dependencies and the size of deployment packages.

For example, a quantum random number function that receives a user's input (such as range or number of qubits) and generates an output with a truly random number can be considered a suitable case. However, there are better candidates for development as a quantum function than a massive quantum machine learning application.

A **quantum service** is the corresponding service of the quantum function deployment, with the API endpoint for invoking. In QFaaS, the service APIs endpoint has the URL endpoint as follows: `<IP or domain name>/api/function/<function-name>` Access to this function is managed by QFaaS Core APIs and can be adjusted by the function owners or the administrator users of the framework. We use the `OAuth2` protocol to manage the access token so authorized users can invoke functions.

## *1.4.2 Sample quantum function development*

Users can develop and test the quantum circuit locally before coding a quantum function compatible with the QFaaS framework. To ensure a smooth operation during function development, deployment, and invocation, we have defined two standardized formats: one for the JSON request and another for the function code structure.

The sample structure of a quantum function in QFaaS is defined as follows:

```
from qfaas import Backend, RequestData, Utils
# Define SDK name
sdk = "<sdk name, such as qiskit, cirq, braket, or qsharp>"

# Pre-processing input data
def pre_process(input):
    [define the data preprocessing]
    return preprocessed_data

# Generate Quantum Circuit
```

```
def generate_circuit(input):
    [define the quantum circuit code based on input]
    return circuit

# Post-processing output data
def post_process(job):
    [define post-processing process]
    return output

def handle(event, context):
    # Pre-processing
    requestData = pre_process(event)
    # Jump to the post processing step if postProcessOnly is set to True
    if requestData.postProcessOnly:
        job = post_process(requestData)
    else:
        # Generate Quantum Circuit
        qc = generate_circuit(requestData.input)

        # Verify and get Backend information
        backend = Backend(requestData, qc)

        # Submit job and wait up to 1 min for job to complete.
        job = backend.submit_job(qc)
        # Post-process
        if job.jobResult:
            job = post_process(job)

    # Generate response data and return to user
    response = Utils.generate_response(job)
    return response
```

1. Import `qfaas` library, a supplement Python library we developed to facilitate the interaction with other components in the QFaaS framework.

2. Define the pre-processing steps for input data of the invocation in the `pre_process` function.

3. Generate the corresponding quantum circuit based on the pre-processed input data in the `generate_circuit` function. This circuit should be adapted to the input data to benefit from the QFaaS approach.

4. Verify and retrieve backend information based on the user's request. Suppose the user manually selects a specific backend for the execution. In that case, that backend must be verified before execution to ensure that 1) the user has permission to use that backend and 2) the quantum backend is valid and operational. The Backend object of the `qfaas` library supports this process.

5. Define the `submit_job` function for submitting the corresponding quantum job to the verified backend. This can also be handled by calling the `backend.submit_job` function in the `qfaas` library

6. Define the post-processing steps for further processing the output after the function is finished processing in the `post_process` function. By default, the maximum waiting time to get the response in an invocation
7. Generate the response and return it to the user in JSON format. The QFaaS library's `Utils` library supports this.

As a sample function, we consider a simple quantum random number generation circuit with a variable number of qubits as follows (see Figure 6):

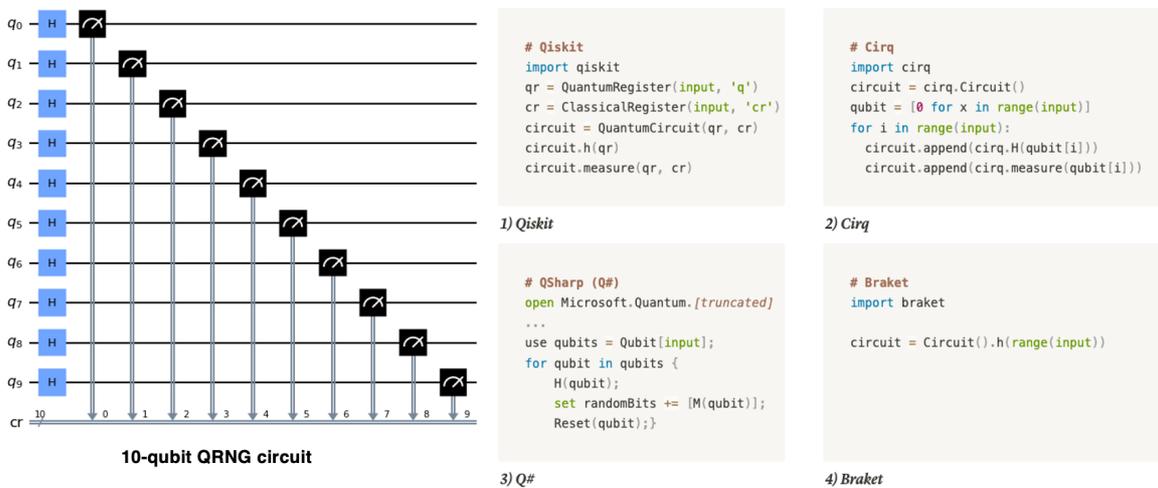

Figure 6. Example for code snippets for generating a simple QRNG function in different SDKs
(Reused with permission from Nguyen et al. [1])

We can define the `generate_circuit(input)` function as in the sample code above. The circuit with different numbers of qubits and gates will be dynamically generated based on the input value from the user's invocation.

### *1.4.3 Deploying quantum functions as services using QFaaS*

After finishing the development of a quantum function in QFaaS, we can deploy it as a service with the support of QFaaS APIs and the CI/CD Automation component. Figure 7 shows the overview of the steps involved in the deployment of functions within QFaaS.

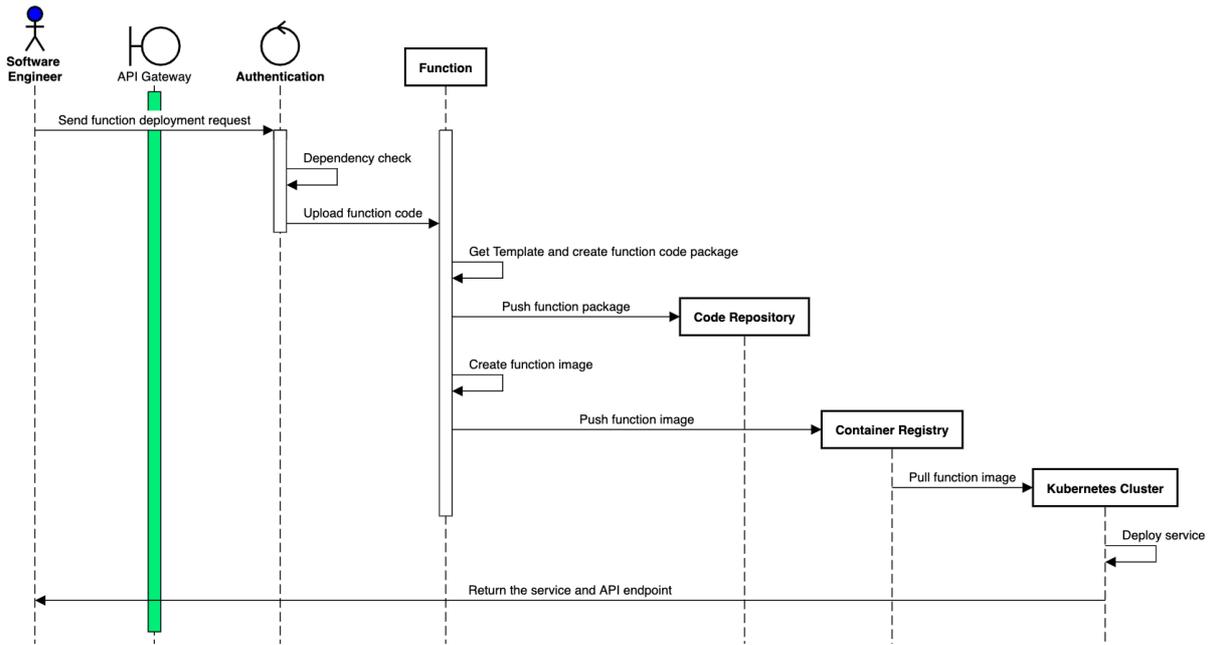

*Figure 7. The sequence of steps in the function deployment in QFaaS*

We provided two main options for function deploying: using the QFaaS UI or directly using QFaaS APIs.

**Option 1: Using QFaaS UI**

We can go to the Functions page, then click **New Function** (see Figure 8)

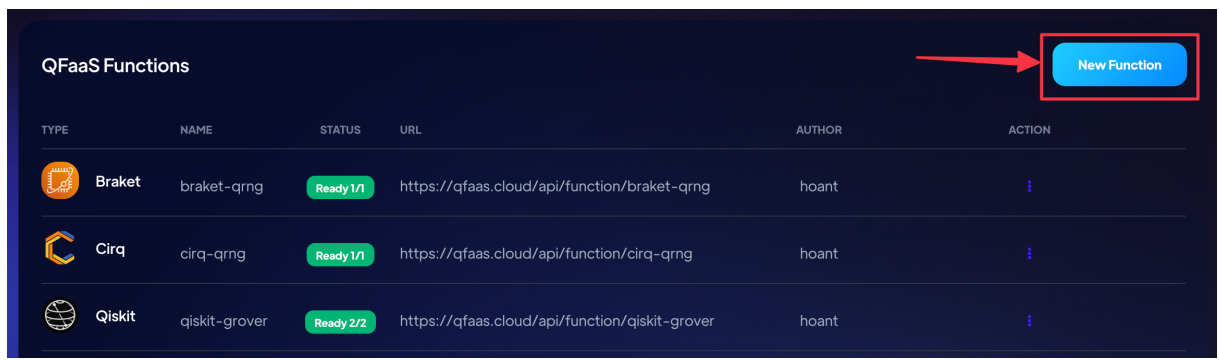

*Figure 8. Create a new quantum function in QFaaS*

Next, we must define the function name, select the function template (SDK), and add the function source code (see Figure 9).

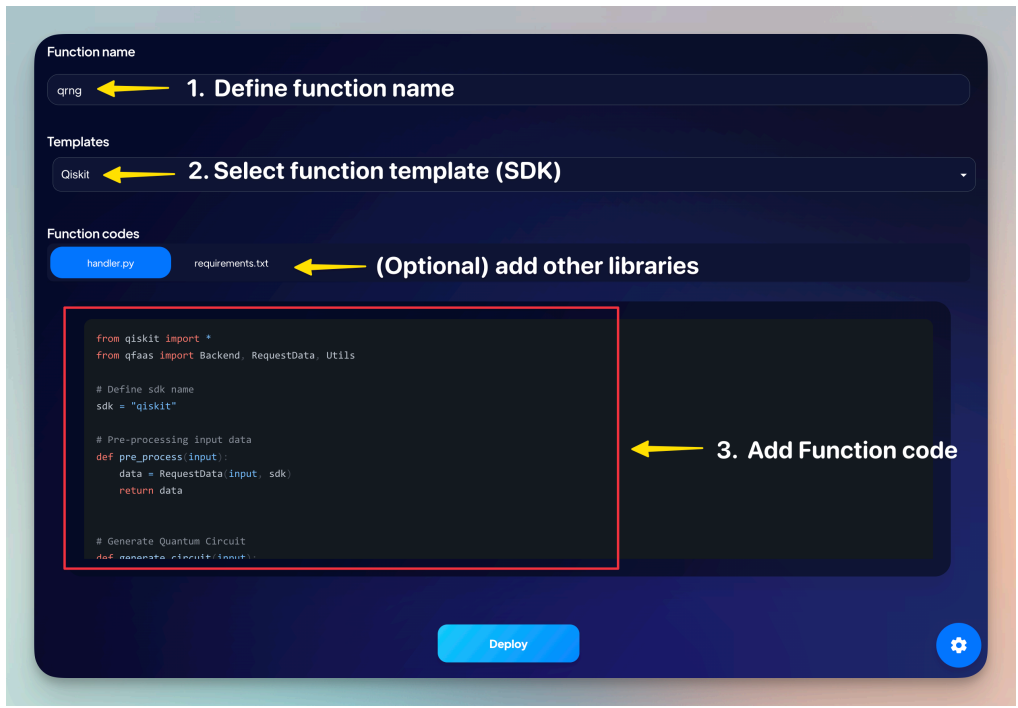

*Figure 9. Sample of new quantum function development page*

1. Define the function name
2. Select a function template that provides a sample function code structure for different SDKs. We support four popular quantum SDKs: Qiskit, Cirq, Q#, and Braket. Each function in QFaaS has a unique identifier: `<template name>-<function name>`. For example, a Qiskit function with the name `qrng` will have the identifier `qiskit-qrng`.
3. Add the function code, following the standard function template provided in the `handler.py` code block. If any additional external Python libraries are needed, specify them in the `requirement.txt` tab.
4. Select Deploy after finishing all the processes above.

The corresponding function will be initialized and automatically deployed to the Kubernetes cluster, following the CI/CD pipeline defined in Gitlab.

**Option 2: Using QFaaS APIs**

We can directly use the function APIs provided by QFaaS Core APIs to deploy a function. This option is recommended for advanced users who want to integrate the QFaaS APIs into their customized workflow. Generally, this is the critical process under the hood of using the QFaaS UI above.

Before using any QFaaS Core APIs, the user needs to have a valid authorization token obtained after successfully logging in with a valid username and password through the QFaaS Login API at `<IP address or domain name>/api/auth/login`.

Each API in QFaaS requires a specific JSON schema that we need to follow. In our case of deploying a function, we need to provide a valid request following the following schema (see Figure 10):

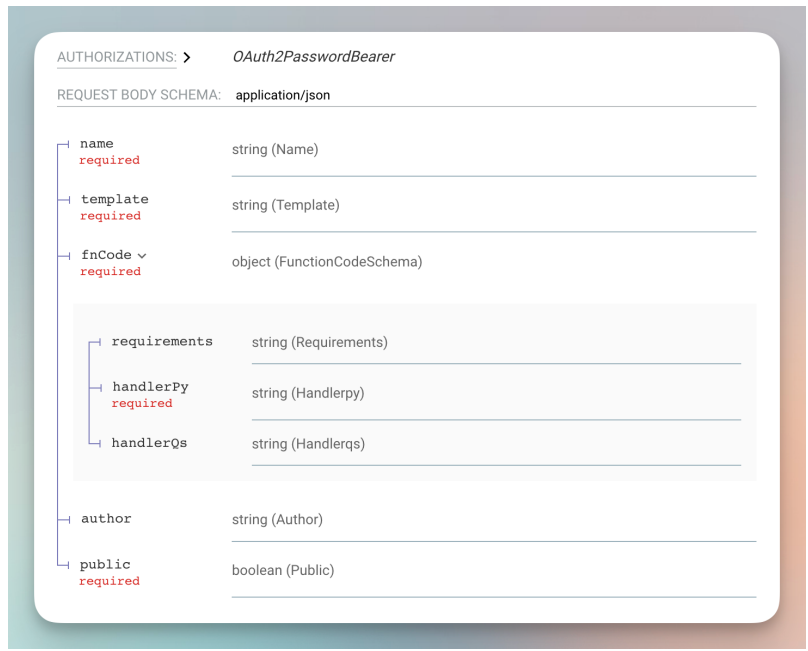

Figure 10. Structure of new quantum function development request

- `name`: define the function name
- `template`: define the function template
- `fnCode`: add corresponding function code objects in Base64 format, including:
    - `requirements`: external library
    - `handlerPy`: main function code
    - `handlerQs`: main function code, only used for Q# function
- `author`: owner of the function
- `public`: make the function publicly visible or keep it private for local testing.

A sample function code request is as follows:

```
{
  "name": "qrng",
  "template": "qiskit",
  "fnCode": {
    "requirements": "<Base64 encoded data of requirements.txt file>",
    "handlerPy": "<Base64 encoded data of function code>",
    "handlerQs": "<Base64 encoded data of Q# function code"
  },
```

```
    "public": true
}
```

After submitting the API request with a valid format and authorization, the function deployment procedure will be triggered, and the corresponding function will be deployed automatically, similar to using QFaaS UI.

After successfully deploying a function, a corresponding service with the API entry point for further invocation will be created. If we use QFaaS UI, that function will be available on the QFaaS dashboard for interaction, as shown in Figure 11.

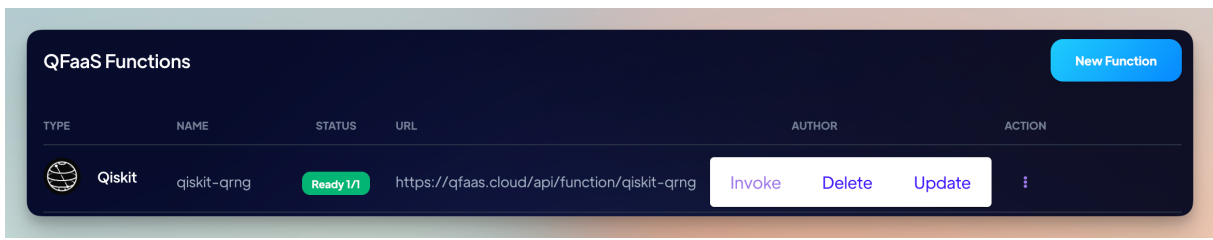

*Figure 11. Sample of a quantum function after its deployment*

## 1.4.4 Invoking and Using Quantum Service in QFaaS

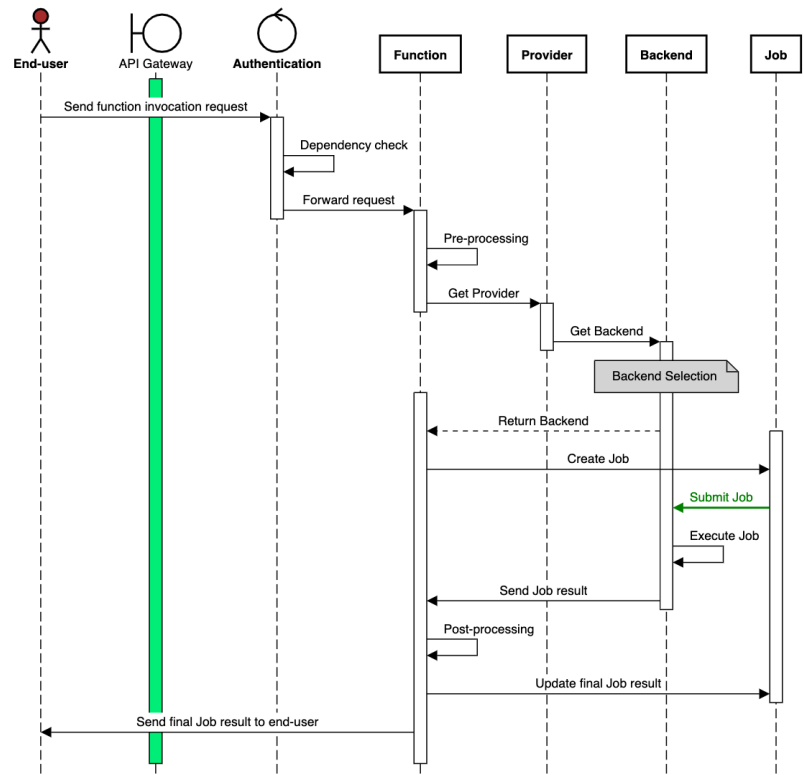

*Figure 12. The sequence of steps in the service invocation of QFaaS*

The corresponding API endpoint will be exposed after a quantum function is deployed as a service on QFaaS. Figure 12 illustrates the sequence of different steps involved in invoking quantum services on QFaaS, including dependency check, pre-processing, backend selection, job submission, and post-processing. The framework automatically handles this invocation process after receiving incoming requests from users.

We can integrate that API endpoint of a quantum service into other applications or invoke the API directly. There are two main ways to invoke quantum services:

1. **Directly using the QFaaS UI**

    Users can invoke the quantum service after its deployment using the QFaaS UI. This serves as a user-friendly interface for testing the quantum function before integrating it into other applications. For example, Figure 13 illustrates invoking the `braket-qrng` function to generate a 20-qubit truly random number. The dashboard allows users to select various execution parameters, such as the backend provider, number of shots, and input data.

    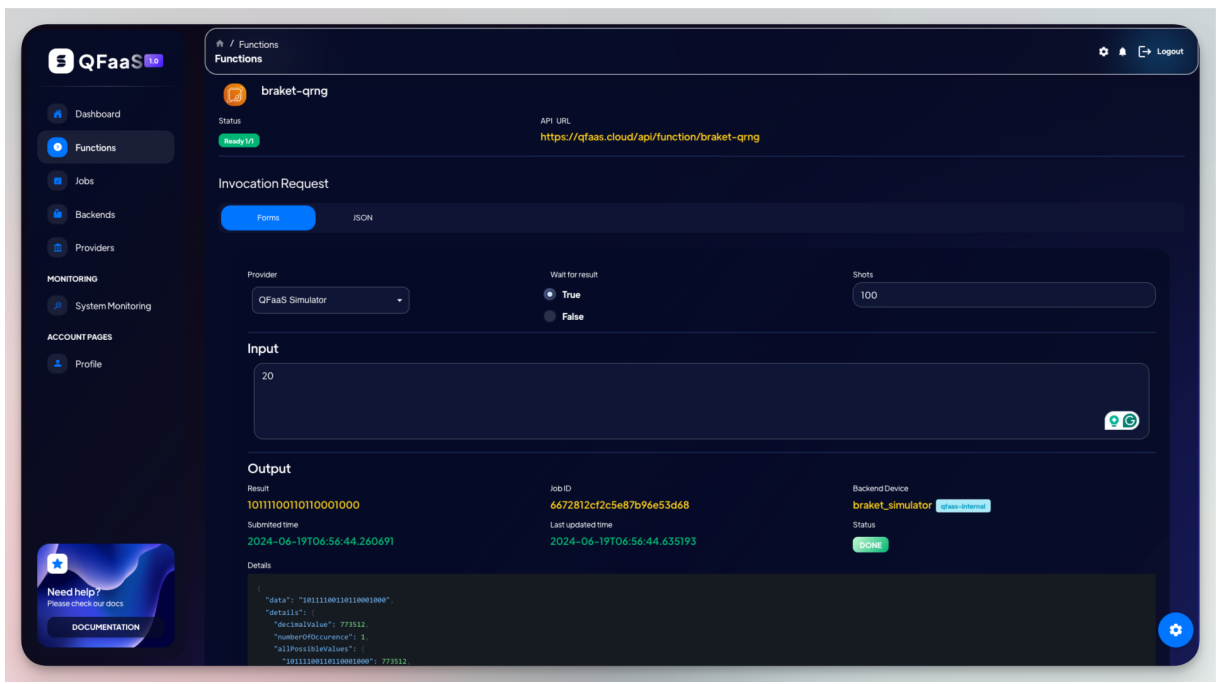

    *Figure 13. Example of invoking the QRNG functions using Braket SDK*

    Another example is shown in Figure 14, which demonstrates factoring numbers into prime factors using Shor's algorithm. For demonstration purposes, we factorize 15 into two prime factors, 3 and 5, using the corresponding Qiskit quantum circuit. Users can customize input and output data formats to adapt to different function requirements.

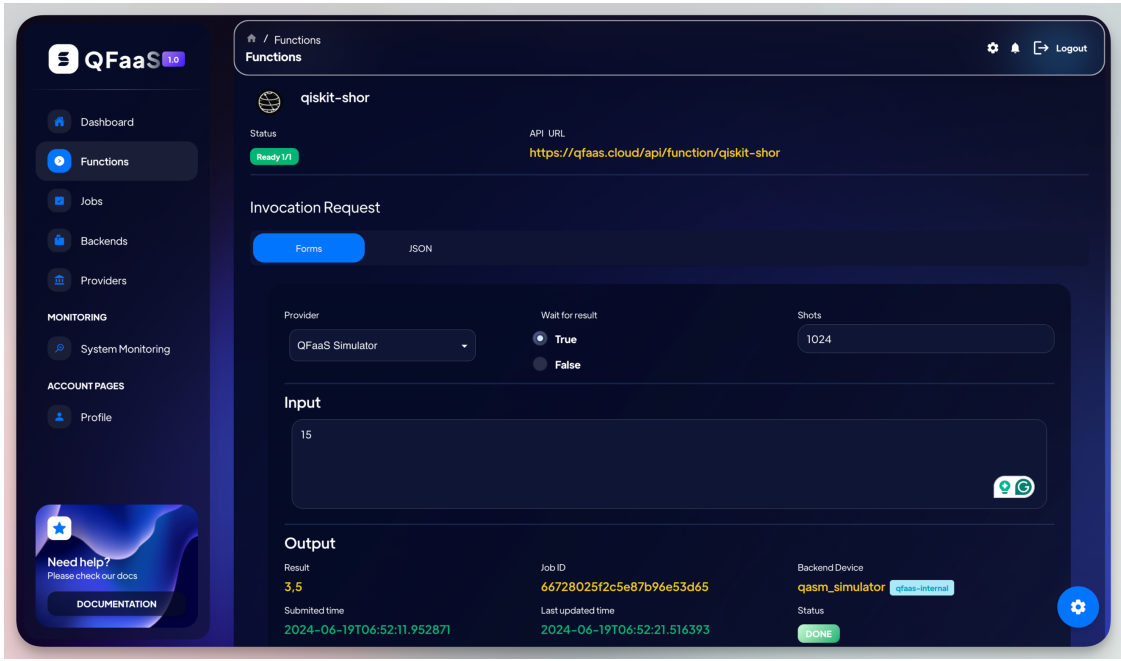

*Figure 14. Example of qiskit-shor quantum function for prime factorization*

**2. Integrating quantum API endpoint into other applications**

Each deployed quantum function (or quantum service) in QFaaS will be associated with an API endpoint, which can be integrated into other existing applications to solve specific tasks with quantum solutions. The default pattern of an API endpoint in QFaaS is: `<domain name>/api/function/<function name>`. For standardization, we define standard JSON input and output formats for all requests and responses of the QFaaS quantum service. A sample function invocation request in JSON format is as follows:

```
{
    "input": <input data>,
    "shots": <number of shots (execution repetition)>,
    "waitForResult": <waiting until getting the result or not>,
    "provider": <provider name>,
    "autoSelect": <auto selection of backend option>,
    "backendType": <type of prefered backend, "qpu" or "simulator">,
    "backendName": <manual specify backend name, or leave blank for auto>
}
```

The invocation response format follows the JSON format below:

```
{
  "data": <post-processed data as output>,
  "details": {<raw data, defined by function developers>}
}
```

These data formats are customizable to ensure flexibility for future framework extensions.

## *1.4.5 Manage Execution Results*

A job in QFaaS is an instance of an invoked execution request to a quantum service identified by a unique ID. Job information and execution results are stored in the QFaaS database for further retrieval and analysis.

The maximum waiting time for a job execution can be defined in QFaaS, as the actual waiting time and completion time at the external quantum provider can vary significantly. By default, the waiting time threshold is set to 1 minute. After the waiting threshold, the user can retrieve the result using the issued Job ID. This setting is essential for jobs sent out for execution in an external quantum cloud provider because the queuing time may take longer than expected. Users can retrieve job information and results later using the Job dashboard in the QFaaS UI or the QFaaS Core API's corresponding API. For example, the job execution result of a QRNG function at the IBM Montreal quantum node is shown in Figure 15.

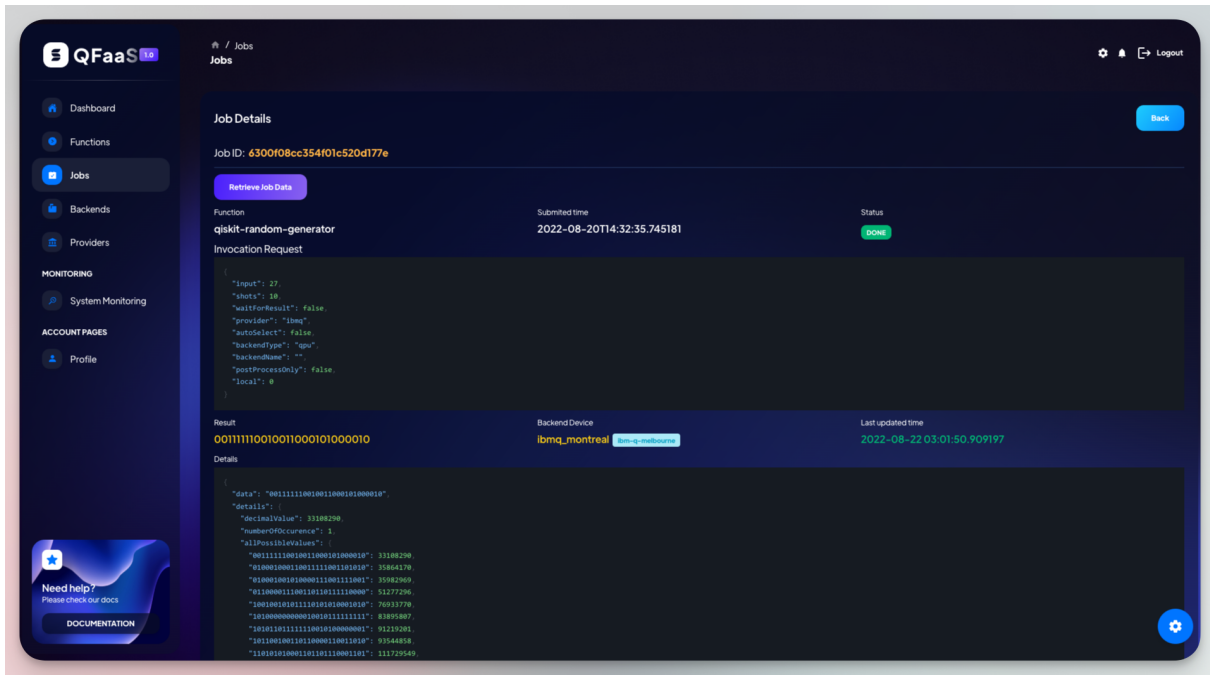

*Figure 15. Sample of a quantum job data retrieval from the QFaaS framework*

## **1.5 Discussion and Outlook**

Serverless quantum computing is the emerging paradigm for quantum computing. However, it is in its infancy, and several key challenges must be addressed. First, a quantum circuit knitting technique [18] should be developed to effectively cut large-scale circuits and distribute them across multiple NISQ quantum nodes for execution. The execution results need to be aggregated to fulfill the expectation of executing the original circuit. This challenge also requires quantum parallel processing and quantum distributed computing [19] techniques, which accelerate the execution of large-scale quantum applications. Current quantum systems only allow a single quantum circuit to be executed at once to mitigate the error caused by crosstalk and decoherence of qubits in the NISQ quantum systems [20]. Second, an efficient hybrid quantum-classical task orchestration [21] is required as many advanced quantum applications usually require the combination of classical optimization tasks and quantum executions, such as quantum machine learning (QML) [22] and variational quantum eigensolver (VQE) algorithms [23]. Machine learning and other state-of-the-art techniques can efficiently address these quantum resource management challenges. A standardized quantum software development workflow [9] is also needed for the diverse SDKs, frameworks, and quantum cloud platforms.

The current development of QFaaS aims to demonstrate a viable approach for realizing the serverless quantum computing paradigm. However, we also recognized several challenges and limitations for continuous improvement in future extension and adaptation. Notably, the early stage of quantum software development kits (SDKs) requires fast adaptation and alignment with the latest updates of all quantum software dependencies. For example, the recent release of Qiskit 1.0 [24], which includes significant updates to its core features, is not backward compatible with previous versions. This necessitates the migration of software and frameworks that utilize earlier versions. Second, limitations of current quantum cloud resources hinder the empirical evaluation of practical quantum applications. Each quantum job must be placed in the waiting queue long before execution. This pattern violated the typical pattern of short-running serverless functions. However, these challenges can be alleviated with future development and the public availability of quantum cloud resources. Finally, additional efforts are required to enhance further the QFaaS features, such as backend selection and quantum cold start mitigation [1]. As a part of the open-source quantum software ecosystem, community contributions to the QFaaS framework are expected to promote its adoption and extension in the future.

## 1.6 Summary

This chapter discusses the serverless quantum computing paradigm and service-oriented quantum applications development using the QFaaS framework. We highlighted the potential of this paradigm for quantum software engineering utilizing current quantum hardware and software development kits. Detailed instructions for deploying and setting up the QFaaS framework were also provided, along with demonstrations on how to use QFaaS to deploy service-oriented quantum functions. This framework serves as a foundation point that can be further extended to support the practical quantum utility era in the future.

**Acknowledgements**

This work is partially supported by the University of Melbourne, Australia, by establishing an IBM Quantum Network Hub and the Nectar Research Cloud. We appreciate the support from the Strangeworks Backstage Program in providing access to Amazon Braket. Hoa Nguyen acknowledges the support from the Science and Technology Scholarship Program for Overseas Study for Master's and Doctoral Degrees, Vingroup, Vietnam.